\documentclass[12pt]{iopart}
\usepackage{natbib, aas_macros}
\usepackage{graphicx}
\usepackage{longtable}
\usepackage{lscape}
\usepackage[usenames, dvipsnames]{color}
\usepackage{txfonts}

\begin{document} 

\newcommand{\gguide}{{\it Preparing graphics for IOP Publishing journals}}

\title{Survival of exomoons around exoplanets}

\author{V. Dobos$^{1,2,3}$, S. Charnoz$^4$, A. P\'al$^{2}$, A. Roque-Bernard$^4$ and Gy. M. Szab\'o$^{3,5}$}

\address{$^1$ Kapteyn Astronomical Institute, University of Groningen, 9747 AD, Landleven 12, Groningen, The Netherlands}
\address{$^2$ Konkoly Thege Mikl\'os Astronomical Institute, Research Centre for Astronomy and Earth Sciences, E\"otv\"os Lor\'and Research Network (ELKH), 1121, Konkoly Thege Mikl\'os \'ut 15-17, Budapest, Hungary}   
\address{$^3$ MTA-ELTE Exoplanet Research Group, 9700, Szent Imre h. u. 112, Szombathely, Hungary}
\address{$^4$ Universit\'e de Paris, Institut de Physique du Globe de Paris, CNRS, F-75005 Paris, France}
\address{$^5$ ELTE E\"otv\"os Lor\'and University, Gothard Astrophysical Observatory, Szombathely, Szent Imre h. u. 112, Hungary}

\ead{vera.dobos@rug.nl}
\vspace{10pt}
\begin{indented}
\item[]January 2020
\end{indented}           


 
\begin{abstract}
Despite numerous attempts, no exomoon has firmly been confirmed to date. New missions like CHEOPS aim to characterize previously detected exoplanets, and potentially to discover exomoons. In order to optimize search strategies, we need to determine those planets which are the most likely to host moons. 

We investigate the tidal evolution of hypothetical moon orbits in systems consisting of a star, one planet and one test moon. We study a few specific cases with ten billion years integration time where the evolution of moon orbits follows one of these three scenarios: (1)~``locking'', in which the moon has a stable orbit on a long time scale ($\gtrsim 10^9$~years); (2)~``escape scenario'' where the moon leaves the planet's gravitational domain; and (3)~``disruption scenario'', in which the moon migrates inwards until it reaches the Roche lobe and becomes disrupted by strong tidal forces.

Applying the model to real cases from an exoplanet catalogue, we study the long-term stability of moon orbits around known exoplanets. 
We calculate the survival rate which is the fraction of the investigated cases when the moon survived around the planet for the full integration time (which is the age of the star, or if not known, then the age of the Sun).The most important factor determining the long term survival of an exomoon is the orbital period of the planet. For the majority of the close-in planets ($<10$~days orbital periods) there is no stable orbit for moons. Between 10 and 300~days we find a transition in survival rate from about zero to 70\%.

Our results give a possible explanation to the lack of successful exomoon discoveries for close-in planets. Tidal instability causes moons to escape or being tidally disrupted around close-in planets which are mostly favoured by current detection techniques.
\end{abstract}

\vspace{2pc}

\maketitle
   


\section{Introduction}
Regular moons are predicted to form generally in planetary systems, as a direct outcome of planet formation. During the core-accretion, gravitational perturbations between planet embryos imply a series of constructive impacts up to the formation of a fully grown planet \citep{nagasawa07}. This phase may happen between a few Myrs to a few 100~Myrs after the star formation \citep[see e.g.][]{chambers98}, and in these processes, satellites may form around the growing planets. Such giant impacts are advocated for the formation of the Earth's Moon \citep{canup04} and for the formation of Uranus and Neptune's satellites \citep{morbidelli12}.  Satellites may also be formed in rings, which could be natural outcomes of giant impacts or tidal disruptions, either in the planet formation phase or later in a relaxed planetary system \citep[see e.g.][]{charnoz09, canup12, crida12}.

An alternative scenario invokes the accretion of moons in the gaseous circumplanetary envelopes that surround the most massive giant planets during their growth in the gaseous protoplanetary disks \citep[see e.g.][]{canup06, sasaki10, miguel16}. Inside the circumplanetary disk, the solid material is replenished by the surrounding protoplanetary disk. This solid material is thought to accrete in the form of large satellites in a way similar to planets. Then the young satellites migrate inward, and sometimes sink into the planet. When the circumplanetary disk disappears, the satellite system remains and then undergo long-term tidal evolution.

After the planets have {been} formed, the orbits of moons evolve due to the planet's tides. Tidal forces lead to the dissipation of energy. On the one hand, this usually results in the gradual expansion of the moon's orbits, if the evolution starts outside the planet's synchronous radius. (A synchronous orbit is where the moon's orbital period is equal to the planet's rotational period.) On the other hand, moons inside the synchronous orbit will migrate inwards and can reach the planet's Roche limit where the moons are disrupted by tidal forces.

The timescale of orbital evolution depends mainly on the intensity of dissipation in the planet's interior, which relies on the material composition  and internal structure of the planet. The dissipation coefficient also varies with the changing planet radius (for gas-rich planets 
Recent papers attempting to compute the quality factor, $Q$, (also known as \textit{tidal dissipation}) show that $Q$ could even be frequency dependant, and thus may strongly depend on the orbital distance \citep[see e.g.][]{ogilvie04}. However, in the current stage of our knowledge, the large uncertainties in the origin of the dissipation process and in the internal structure of detected exoplanets preclude any theoretical calculation of $Q$ for a specific planet. For this reason, in the present paper, a classical approach is adopted where $Q$ will be considered as a constant, and many values will be explored. 
The dissipation scales as $Q^{-1}$ and we can assume it to be in steady-state when exploring those scenarios which lead to moons on the long-term stability. For rocky planets, $Q$ is low and the dissipation is high ($Q \sim$10--100), conversely for giant planets $Q$ is high ($\sim 10^4$--$10^5$) and dissipation is low \citep{goldreich66, alvarado17}. 

The quest for exomoon discovery is a challenging task in itself, and upon finding an exomoon we could empirically constrain the $Q$ value of the host planet by considering the planetary distance and the age of the star. This will provide invaluable information about the exoplanet interior, too, since high/low dissipation rates may depend on the planet core and interface with mantle \citep{lainey12,remus12}.

Recently, \citet{alvarado17} investigated the orbital changes of exomoons around close-in giant planets due to tidal effects. In addition to the tidal evolution, they implemented the changes of the giant planet's physical parameters into the model. The radius of the planet decreases over time, and  this change is most prominent in the first billion year after the formation of the planet. The value of $k_2/Q$ slowly increases, which also has an impact on the orbital evolution of the moon. 

Probably the most intriguing question concerning the moons is the lack of a firmly confirmed moon in exoplanetary systems. Before novel observation projects are designed for exomoon detection, first we need to find the answer to the question where the exomoons would be. Did we use suboptimal signal detection strategies, or did we simply look for moons in systems where their existence is not too likely? This latter possibility is the central question of this paper.

\citet{barnes02} showed that exomoons are more stable at longer planetary orbital periods, and also that the orbit of exomoons can tidally evolve first outward and then inward (towards the planet) as the planetary rotation slows due to tidal interaction with the star. \citet{heller12} studied the orbital stability of habitable moons using the method of \citet{kipping09} and it was found that for stars of masses below $\approx 0.2 M_\odot$, moons are not likely to stay on stable orbits around planets which are in the habitable zone. Later \citet{zollinger17} showed that for planets in the habitable zones of $\lesssim 0.5 M_\odot$ stars, the orbit of moons are not stable as they migrate towards the planet.

Here we investigate the evolution of hypothetical satellites orbiting the currently known exoplanets with numerical integration, under the influence of tidal interaction. We map a wide range of parameters, including planet periods, moon to planet mass ratios, and $k_2/Q$.  Instead of studying a few cases in large details and with high accuracy, which would be meaning less due to our lack of knowledge on both the physical origin of tidal dissipation, and on the planet’s internal structure, we are more interested in the statistical survival rate of moons. In order to investigate a large range of configurations over billions of years of evolution, we used the average description of \citet{barnes02} which are formulated in Eqs. 1--3 (Section \ref{Methods}). We apply this method to identify different evolution pathways of moons (Section \ref{examples}) and then we calculate survival rates of hypothetical moons around known exoplanets with the aim of pointing out to specific planets as observation targets which might have satellites (Section \ref{survival} and \ref{table}).
 
Independently of our work, \citet{sucerquia19} used similar methods to investigate the orbit stability of exomoons around close-in planets, and estimated their detectability through TDV and TTV effects. \citet{guimaraes18} had a similar goal with the focus on the photometric detectability of hypothetical moons around Kepler planets. They selected 54 planets which could possibly host observable moons. Although we also list planets with survival rates for moons, we focus our investigations on the survival statistics.

\section{Methods} \label{Methods}

\subsection{Orbital evolution of satellites}

We calculate the orbital evolution of the planet and the moon from the following equations \citep{barnes02, sasaki12}.


\begin{equation}
\frac{\mathrm{d}n_\mathrm{m}} {\mathrm{d}t} = - \frac{9} {2} \frac{k_{2 \mathrm{p}} R_\mathrm{p}^5 G M_\mathrm{m} n_\mathrm{m}^{16/3}} {Q_\mathrm{p} (G M_\mathrm{p})^{8/3}} \mathrm{sgn}(\Omega_\mathrm{p} - n_\mathrm{m}) \, ,
\label{dnmperdt1}
\end{equation}

\begin{equation}
\frac{\mathrm{d}n_\mathrm{p}} {\mathrm{d}t}= - \frac{9} {2} \frac{k_{2 \mathrm{p}} R_\mathrm{p}^5} {Q_\mathrm{p} G M_\mathrm{p} \left(G M_\star \right)^{2/3} } n_\mathrm{p}^{16/3} \mathrm{sgn}(\Omega_\mathrm{p} - n_\mathrm{p}) \, ,
\label{dnpperdt}
\end{equation}

\begin{equation}
    \frac{d\Omega_\mathrm{p}} {\mathrm{d}t} = - \frac{3} {2} \frac{k_{2 \mathrm{p}} R_\mathrm{p}^3} {Q_\mathrm{p}} \frac{ (G M_\mathrm{m} )^2 } { \alpha (G M_\mathrm{p} )^3 } n_\mathrm{m}^4 \mathrm{sgn}(\Omega_\mathrm{p} - n_\mathrm{m}) - \frac{3} {2} \frac{k_{2 \mathrm{p}} R_\mathrm{p}^3} {Q_\mathrm{p}} \frac{ 1 } { \alpha G M_\mathrm{p} } n_\mathrm{p}^4 \mathrm{sgn}(\Omega_\mathrm{p} - n_\mathrm{p}) \, ,
\label{doperdt}
\end{equation}


\noindent where $n_\mathrm{m}$ and $n_\mathrm{p}$ are the mean motions of the moon and the planet, $k_{2 \mathrm{p}}$ is the second order Love number of the planet which relates to the dilatation due to tidal response, $Q_\mathrm{p}$ is the quality factor of the planet which describes the dissipation of energy in the body, $G$ is the gravitational constant, $R_\mathrm{p}$ is the radius of the planet, $M_\mathrm{m}$, $M_\mathrm{p}$ and $M_\star$ are the masses of the moon, the planet and the host star, respectively, 
$\Omega_\mathrm{p}$ is the rotation frequency of the planet, and $\alpha$ is the prefactor of the moment of inertia ($\alpha M_\mathrm{p} R^2$). 
Circular orbits are assumed for both the planets and the satellites.

Since Equations (1) and (2) have the form of $\dot y=C \, y^p$ (where $y$ is
either $n_{\rm m}$ or $n_{\rm p}$, $p$ is the exponent of $16/3$ while $C$
contains all of the astrophysical constants) and Equation~(3) can directly be
integrated once the time evolution of $n_{\rm m}$ and $n_{\rm p}$ are known,
this set of equations can be solved analytically on the domains until $n_{\rm
m}$, $n_{\rm p}$ or $\Omega_{\rm p}$ do not surpass each other; in this case,
the presence of the ${\rm sgn}()$ function would not allow the fully
analytical solution. We therefore implemented the solution in a
piecewise-analytical manner, monitoring the sign changes in the differences
between $n_{\rm m}$, $n_{\rm p}$ or $\Omega_{\rm p}$ while in the distinct
pieces, the exact soluion is evaluated. Thus, besides the vicinity of the
points where the signs commute, the solution does not even depend on this
stepsize, allowing us to highly speed-up the computations.

For calculating the survival rate of moons around known exoplanets, we let the orbits of the planet and the moon to evolve until the current age of the star, or if it is not know then until the age of the Sun (4.57~Gyr). 
We solve the equations analytically with fixed time steps to monitor whether the moon reaches the Roche lobe of the planet or the critical distance (half of the Hill sphere). In these cases the integration ends because the moon is considered to be lost  
either because of tidal disruption or escape from the planet's 
gravitational bond \citep[for the latter case see][]{domingos06}.
The critical distance is calculated by $R_\mathrm{crit} = 0.5 \, R_\mathrm{Hill} = 0.5 \, a_\mathrm{p} \left( M_\mathrm{p} / (3 M_\star) \right)^{1/3}$.

If at the end of the calculation the moon is still orbiting the planet, then it is counted as a "survival" case.

\subsection{Parameters used for the integrations} \label{order}

Stellar mass ($M_\star$), planet mass ($M_\mathrm{p}$), semi-major axis ($a_\mathrm{p}$) and radius of the planet ($R_\mathrm{p}$) were taken from the The Extrasolar Planets Encyclopaedia (http://exoplanet.eu, data obtained on March 25, 2021). Only those planets are considered where the $M_\mathrm{p}$ or the $M_\mathrm{p} \cdot \sin i$ or the $R_\mathrm{p}$ value is known (at least one of them). Planets with higher masses than 13~Jupiter mass are ignored as well as those with $M_\star < 0.08 M_\odot$. In addition, the selected planets also had to have either the orbital period ($P_\mathrm{p}$) or the semi-major axis data. After this selection 4064 planets remained.

For both the radius and mass (or $M_\mathrm{p} \cdot \sin i$ if $M_\mathrm{p}$ is not given) of the planet, the uncertainties were taken into account wherever possible. If there was no uncertainty in the data, then simply the given mass or radius was used. If there was a symmetric uncertainty given, then the mass or radius values were generated by choosing a random value with a Gauss distribution. The symmetry was determined with the same method as described in \citet[][Section 2.2]{chen17} where the difference between the upper and lower errors must be below 10\%. For these symmetric cases the mean of these two values was taken as the standard deviation of the data. If the difference was higher than 10\%, then the errors were ignored and simply the mean value was used. We applied exactly the same method for those planets where only the $M_\mathrm{p} \cdot \sin i$ value was known, not the real mass. In these cases the error of the $M_\mathrm{p} \cdot \sin i$ was taken into account wherever possible. This way we could generate realistic data for all planets, taken into account the uncertainties in the measurements.

There are 1007 cases (out of 4064) where both the radius and the mass ($M_\mathrm{p}$ or $M_\mathrm{p} \cdot \sin i$) of the planet are known. For all the other planets, the Forecaster model was used for generating realistic values for the missing parameter \citep{chen17}. This tool predicts the mass (or radius) of the planet based on the radius (or mass). As an input, the generated (above-mentioned Gauss distributed) mass (or radius) values were used as posteriors. This way, for each run the planet had a different mass or radius, or both (if the original data had symmetric errors).


See the following description of the defined ranges and distribution of the chosen random values for the rest of the (initial) parameters. 

\begin{itemize}
\item 
\textit{Mass of the moon:} The mass of the hypothetical moons is selected randomly with a uniform distribution between 1\% and 10\% of the host planet's mass, for each planet.
\item 
\textit{Semi-major axis of the moon:} The initial semi-major axis of the moon is chosen randomly with a uniform distribution on a logarithmic scale between 2~$R_\mathrm{p}$ and $R_\mathrm{crit}$.
\item 
\textit{Radius of the moon:} Estimating the radius of exomoons is very challenging, because only the Solar System serves as an example. We choose to define it through their densities ($\rho_\mathrm{m}$) using their mass which is already chosen. Based on the big moons (radius $>$200~km) in the Solar system, we define three categories depending on the locations with respect to the snowline ($a_\mathrm{snow}$). This is because beyond the snowline there are more icy, hence less dense bodies in the Solar system. The snowline is calculated as $a_\mathrm{snow} = T_\mathrm{eff}^2 \cdot R_\star / T_0^2$ where $T_\mathrm{eff}$ is the effective temperature of the star and $T_0 \approx 230$~K is the equilibrium temperature at the planet’s sub-stellar point \citep{cowan11}. A fourth category is also given for those cases when $T_\mathrm{eff}$ or $R_\star$ is not known, hence $a_\mathrm{snow}$ cannot be calculated. The $\rho_\mathrm{m}$ values are selected randomly with a Gauss distribution with the following parameters.
    \begin{itemize}
    \item
    If $a_\mathrm{p} < a_\mathrm{snow}$: the mean value of $\rho_\mathrm{m}$ is 3~g/cm$^3$, $\sigma = 1/3$~g/cm$^3$.
    \item
    If $a_\mathrm{snow} \leq a_\mathrm{p} < 2 a_\mathrm{snow}$: the mean value of $\rho_\mathrm{m}$ is 2.5~g/cm$^3$, $\sigma = 1/3$~g/cm$^3$.
    \item
    If $2 a_\mathrm{snow} \leq a_\mathrm{p}$: the mean value of $\rho_\mathrm{m}$ is 2.5~g/cm$^3$, $\sigma = 1/6$~g/cm$^3$.
    \item
    If $a_\mathrm{snow}$ is not known: the mean value of $\rho_\mathrm{m}$ is 2.5~g/cm$^3$, $\sigma = 1/2$~g/cm$^3$.
    \end{itemize}
\item 
\textit{Rotation period of the planet:} In the Solar system the planet that has the slowest spin around its axis is Jupiter with 9.9~hours. \citet{scholz18} defines an empirical spin--mass relation for planets and brown dwarfs. According to their work, the spin velocity scales as $v \sim \sqrt{M}$, and from this relation it can be concluded that the spin period does not exceed 5~days even for brown dwarfs. (Spin--orbit resonances are not taken into account.) Based on these information and considering possible deviations, we set the spin period to a random value with uniform distribution between 10~hours and 5~days. 
\item 
\textit{Quality factor of the planet:} For the quality factor ($Q_\mathrm{p}$) three categories are definied.
    \begin{itemize}
    \item
    \textit{Rocky planets} if $R_\mathrm{p} < 2 R_\oplus$: $10 < Q_\mathrm{p} < 500$ \citep{goldreich66}. Since lower values are more probable, the distribution of the randomly chosen $Q_\mathrm{p}$ value is uniform on a logarithmic scale.
    \item
    \textit{Ice/gas giants} if $R_\mathrm{p} \geq 2 R_\oplus$ and $P_\mathrm{p} > 10$~days: $10^3 < Q_\mathrm{p} < 10^6$ \citep{goldreich66, ogilvie04, lainey12, lainey17}. For these planets intermediate values are the most probable in this range, hence $Q_\mathrm{p}$ values are chosen with a log-normal distribution with a mean of $Q_\mathrm{p} = 10^{4.5}$ and $\sigma = 10^{0.5}$.
    \item
    \textit{Hot jupiters} if $R_\mathrm{p} \geq 2 R_\oplus$ and $P_\mathrm{p} \leq 10$~days: $Q_\mathrm{p} \sim 5 \cdot 10^6$ \citep{ogilvie04}. We do not have much information on the possible $Q_\mathrm{p}$ value of these planets, hence we use a Gauss distribution with $Q_\mathrm{p} = 5 \cdot 10^6$ as the mean value and $\sigma = 2 \cdot 10^6$.
    \end{itemize}
\end{itemize}

$Q_\mathrm{p}$ is a critical parameter, as it controls the tidal evolution of the planet, however its value is very poorly constrained, and, in the current state of our knowledge it could vary by several order of magnitudes even for similar planets \citep{remus12}.
For the giant planets in the Solar System, $Q$ is mostly constrained by the secular acceleration of a big moon orbiting relatively close to the planet and also by theoretical considerations. For these planets $Q$ is estimated to be in the range of $10^4$--$10^5$ \citep[see e.g.][]{goldreich66, ogilvie04, lainey09}. However, if $Q$ is not treated as a constant but as a function of the tidal frequency, then we obtain much lower values for Saturn with $Q$ about 2500 \citep{lainey12,lainey17}. $Q$ is mostly unknown for ice giants like Uranus and Neptune, but theoretical considerations put them in the range $>10^4$ \citep{goldreich66}. However this estimate is based on some unconstrained assumption on the moon formation scenarios and ages. There is also a more recent estimation for Neptune which puts the $Q$ value between 9000 and 36000 \citep{zhang08}.

Conversely, the $Q$ of terrestrial planets is very low, in comparison to gas giants, and is typically between ten and a few hundreds \citep[see e.g.][]{macdonald64, goldreich66, smith76, ray01, rainey06, lainey07}. The $Q$ for these planets is better constrained because of the possibility of laboratory tests of rocks (where possible). While $Q$ might go up to 500 for rocky planets, it is typically below 200 in the Solar System.

It is highly challenging to draw conclusions about the possible $Q$ value of exoplanets. It seems reasonable to assume that rocky planets have similar values as in the Solar System. We can also speculate that maybe the $Q$ of giant planets is in the same order of magnitude as their Solar System counterparts, however, theoretically, with low-density ice-rich cores, $Q$ in the range a few hundreds would also be possible for a Saturn-like planet \citep{lainey17}. In hot-jupiters, because of the assumed synchronous spin and circularized orbit, $Q$ might be around $5 \cdot 10^6$ \citep{ogilvie04}.

The Love number $k_2$ varies modestly from one planet to another in the Solar System but is between 0.1 and 0.6 \citep{gavrilov77}. We choose $k_{2 \mathrm{p}}=0.5$. However $k_2$ never appears alone, but only in $k_2/Q$. So we fix the value of $k_2$ and vary the value of $Q$ to save a free degree of liberty that would lengthen the computation time.

For the same reason, the parameter used for calculating the gyration radius is also fixed to $\alpha = 0.3$. It is expected that planets with different compositions and structures have different gyration radii, but by keeping $\alpha$ fixed, a free parameter is removed.

For each planet ten thousand runs are made to get a wide sample of the different initial parameters.
What we call the "survival rate" is simply the ratio of surviving moons (those that stayed in orbit around the planet until the end of the integration) and the total number of tested configurations. 

For a few example cases, the parameters of the planet and the moon are set to specific values which are described in details in Section \ref{examples}.

\section{Example cases} \label{examples}

In the example cases (see Figs.~\ref{massratio2} and \ref{massratio}), the system configuration consists of a Solar-mass star with an Earth-like planet and a moon which has different mass in the different cases. The quality factor of the planet is 
$Q_\mathrm{p}=10$ in all cases. 
The initial parameters used for the different runs are summarized in Table \ref{initial}. Note that the time shown in the horizontal axis is significantly shorter in Fig. \ref{massratio2} (where it ends at $5 \cdot 10^6$~years) than in Fig. \ref{massratio} (where it ends at $10^{10}$~years).

\subsection{Evolution outcomes}

In the integrations, the moon has been observed to follow three different families of evolutionary tracks, and depending on the actual parameter set, all of these three tracks can evolve rapidly, resulting in an unstable system, or the evolution can be slow enough to lead to a surviving moon. Both the morphology and the stability depends on the position of the Hill sphere, thus eventually, on the masses and orbital periods of both the planet and the moon, the stellar mass, and $Q_\mathrm{p}$.

\begin{table}
\centering
\begin{tabular}{l c c c c c}
\hline\hline
 & Fig.~\ref{massratio2} upper & Fig.~\ref{massratio2} lower & Fig.~\ref{massratio} upper & Fig.~\ref{massratio} middle & Fig.~\ref{massratio} bottom \\
 & panel & panel & panel & panel & panel \\
 \hline
 $M_\mathrm{m}/M_\mathrm{p}$ & 1/85 & 1/85 & 1/85 & 1/150 & 1/300 \\  
 $a_\mathrm{p}$ [AU] & 0.3 & 0.3 & 1.0 & 1.0 & 0.8 \\
 $n_\mathrm{m}$ [10$^{-6}$ s$^{-1}$] & 38.0 & 6.1851 & 2.6640 & 2.6640 & 2.6640 \\
 \hline
\end{tabular}
\caption{Initial parameters used for the example cases shown in the different panels of Figs.~\ref{massratio2} and~\ref{massratio}. The star mass is $M_\star =1  M_\odot$, the planet mass is $M_\mathrm{p} = 1 M_\oplus$ and the quality factor of the planet is $Q_\mathrm{p} = 10$ in each case.}
\label{initial}
\end{table}

\subsubsection{Disruption scenario}

When the planet is close enough to the star, the tidal torque on the planet from the star very effectively spins down the planet. The tidal bulge is shifted slightly toward the trailing side of the moon's orbit, the moon quickly looses energy, spirals inward and is disrupted at the Roche lobe. The time scale of this migration is very short (in the order of 1--200 million years, mostly depending on the stellar distance and on the moon--planet mass ratio) compared to the 10~Gyr integration time, therefore we consider that these moons are unstable. This scenario is described in details in \citet{alvarado17}, and our results are completely compatible with their findings (Fig. \ref{massratio2}, top panel).

\subsubsection{Escape scenario}

\begin{figure}
	\centering{
	\includegraphics[width=24pc]{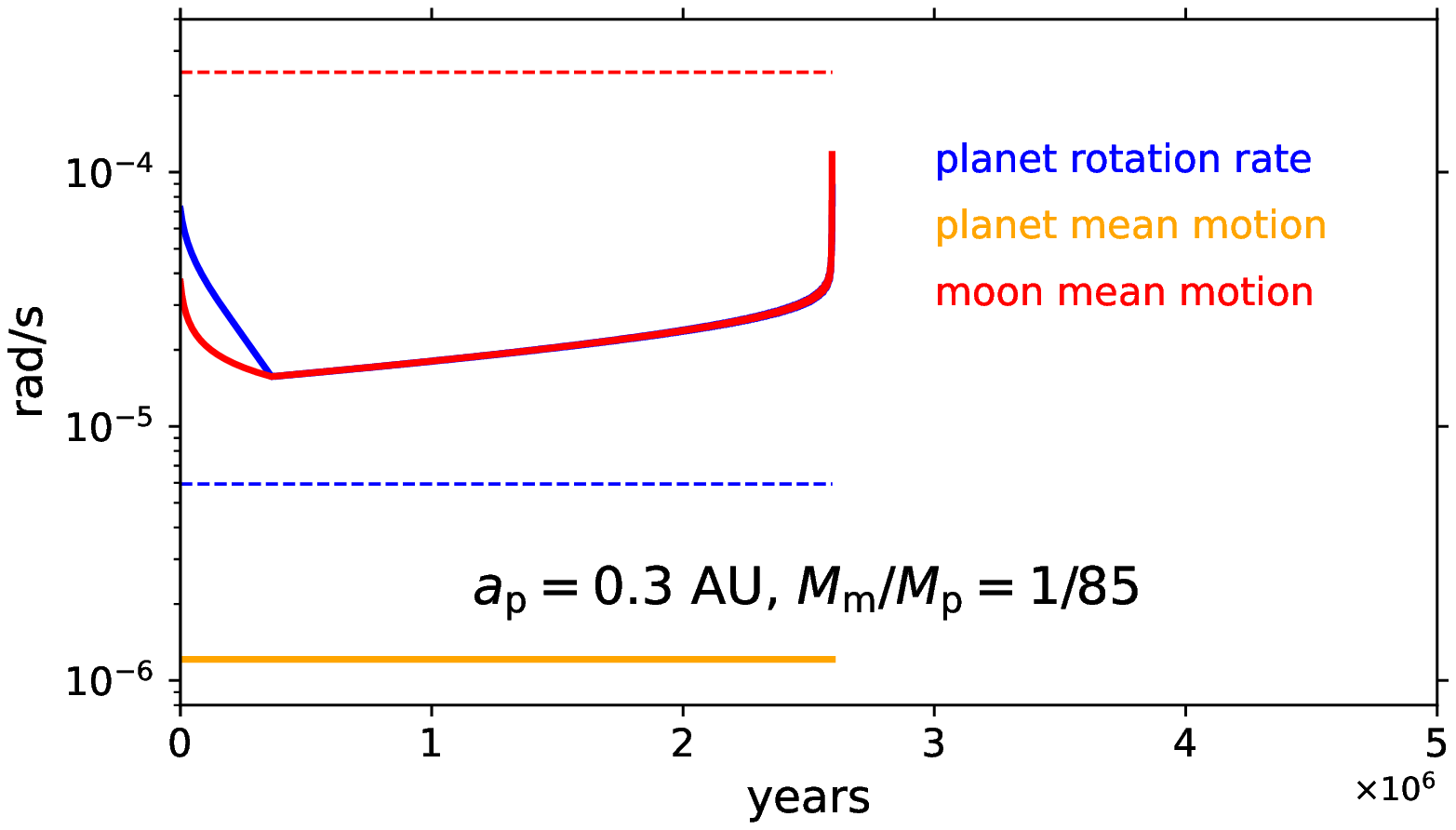}
    \includegraphics[width=24pc]{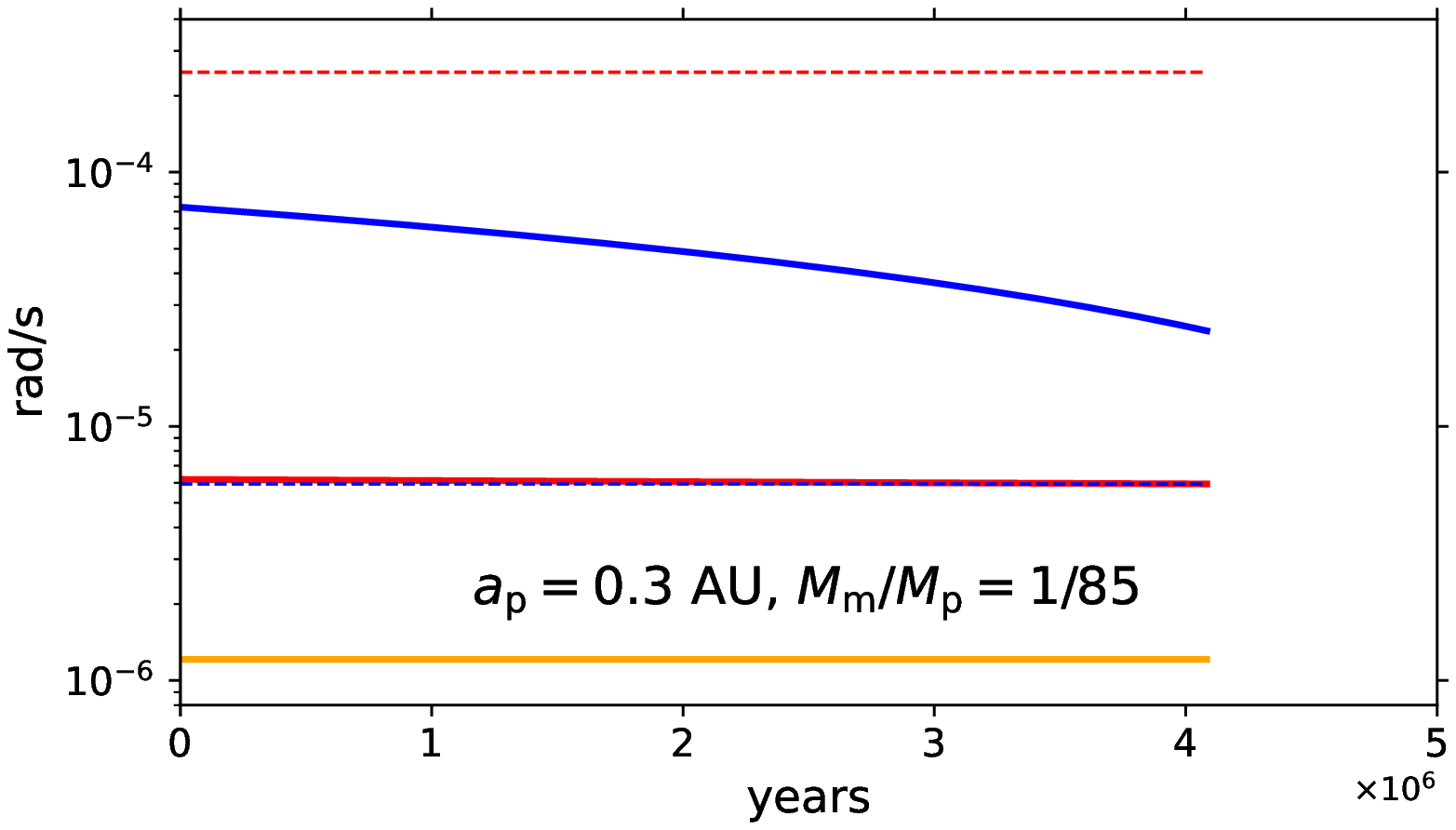}}
\caption{Moon orbit and planet spin evolution in two example cases. See the initial parameters in Table~\ref{initial}. Blue dashed horizontal line at $6 \cdot 10^{-6}$~s$^{-1}$ represents the location of the critical distance from the planet (half of the Hill radius of the planet) beyond which the moon escapes. Red dashed horizontal line shows the Roche radius of the planet expressed in the unit of mean motion. \textit{Upper panel (disruption scenario):} the moon quickly synchronises with the planet, but when migrating inwards to the planet, it reaches the Roche lobe and becomes tidally disrupted. \textit{Lower panel (escape scenario)}: the moon slowly migrates outwards meanwhile the planet spin evolves. When the moon reaches the critical distance (after about 4 million years), it escapes from the planet.}\label{massratio2}
\end{figure}

If the planet's orbit is close to the star, then the Hill radius is smaller. On any orbit inside the Hill sphere, the moon suffers strong tidal forces, especially if the planet is relatively massive, which makes the outward migration very fast. As a consequence, if the moon formed close to the critical distance from the planet (half of the Hill sphere), then it will quickly escape, before synchronization could happen (Fig. \ref{massratio2}, lower panel).

\begin{figure}
	\centering{
	\includegraphics[width=24pc]{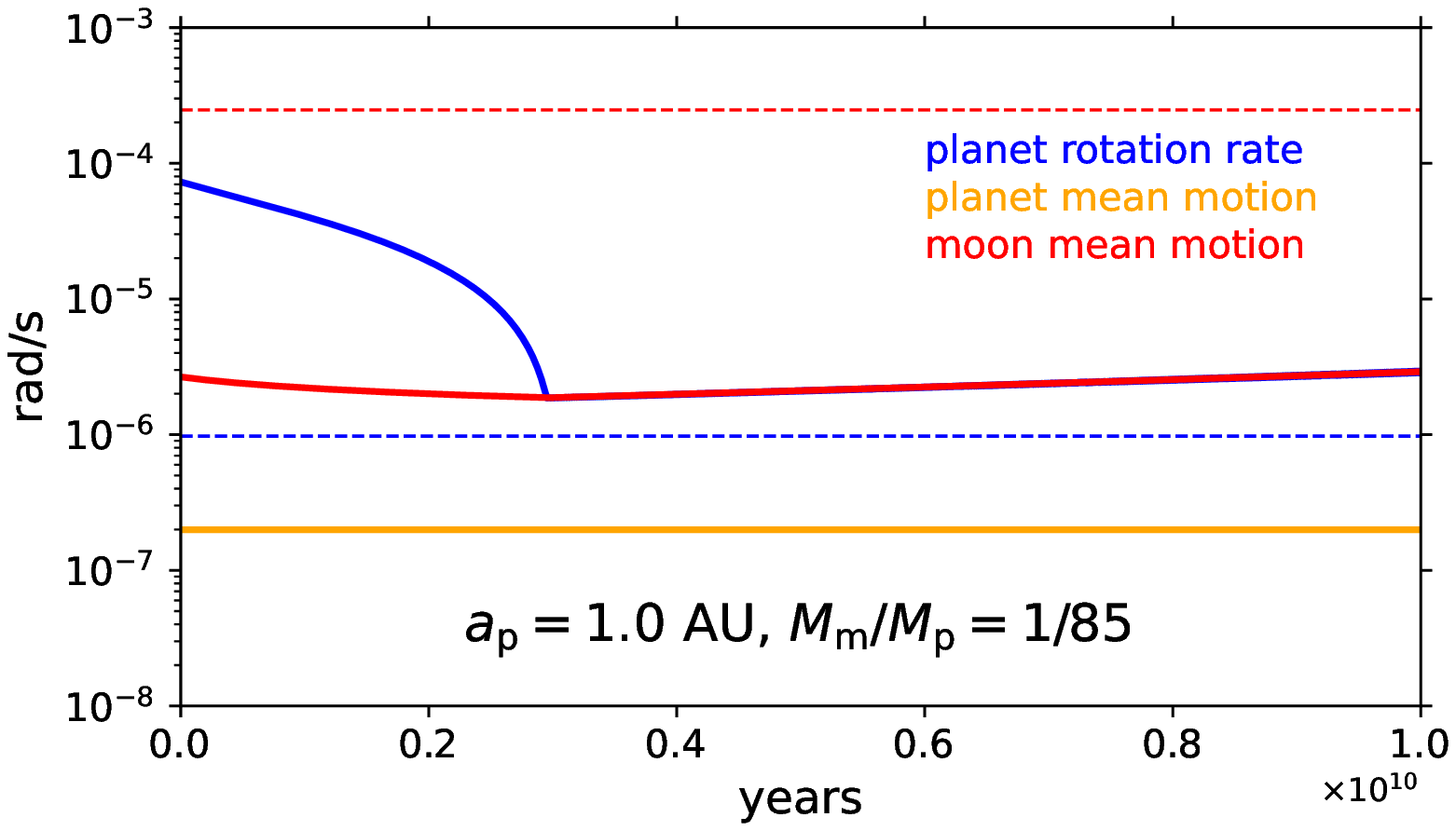}
	\includegraphics[width=24pc]{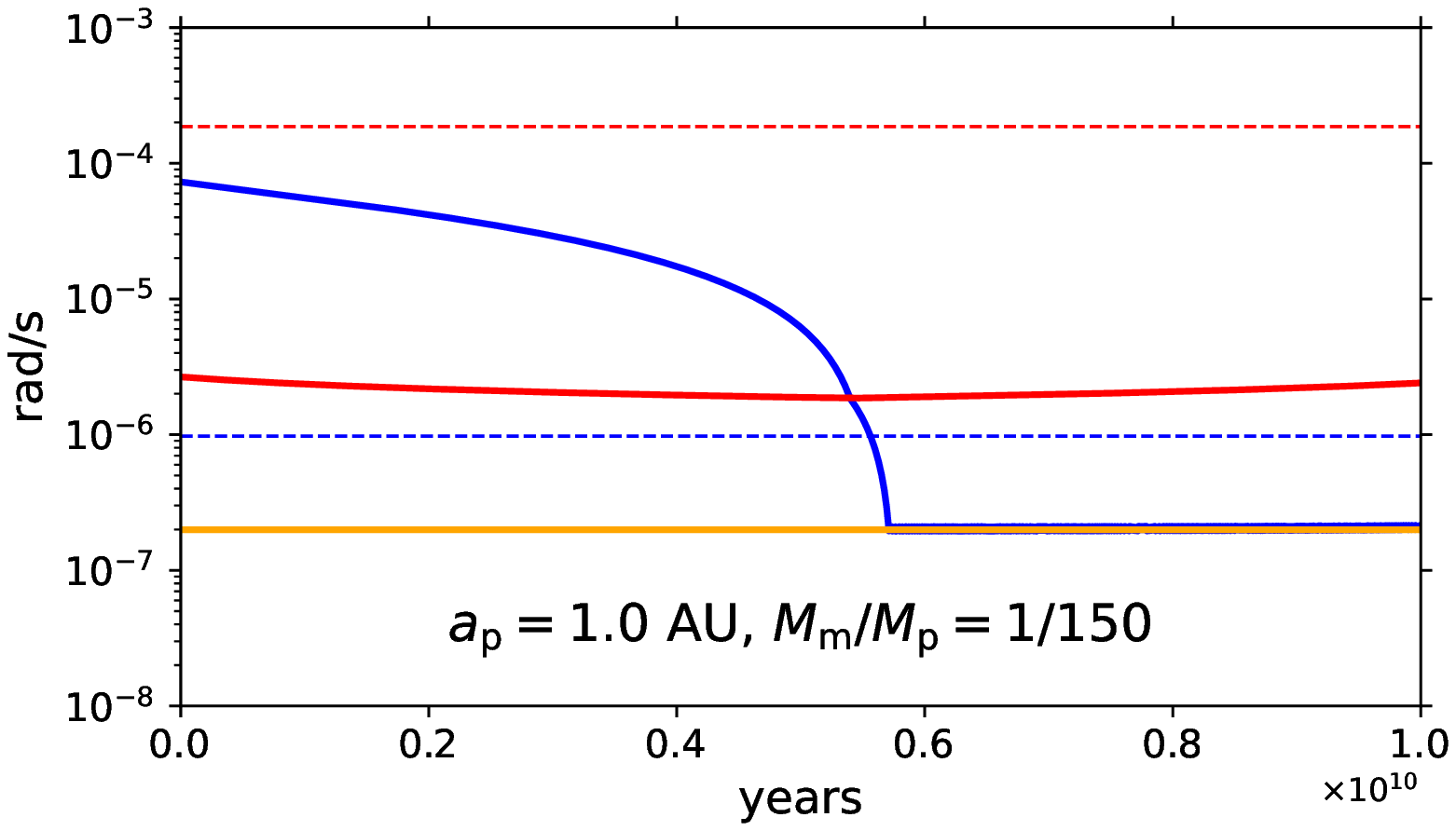}
	\includegraphics[width=24pc]{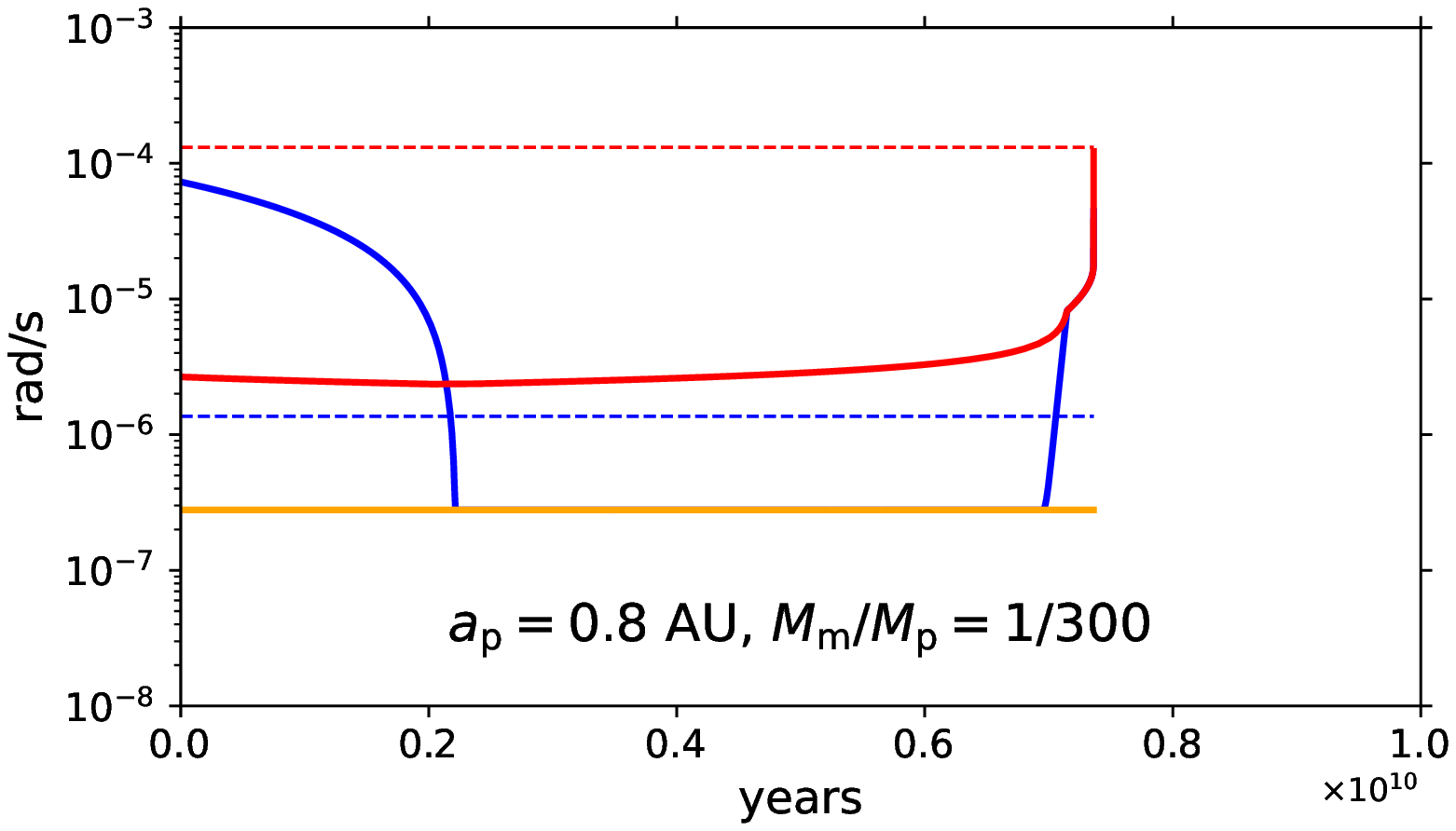}}
\caption{Moon orbit and planet spin evolution in three example cases.  See the initial parameters in Table~\ref{initial}. Blue and red dashed horizontal lines represent the location of the critical distance for moon escape and the location of the Roche limit, respectively, expressed in the unit of mean motion. \textit{Upper panel (moon--planet locking scenario):} the moon synchronises with the planet after $\sim 3 \cdot 10^9$ years; up to this point, the moon slowly migrates outwards from the planet, but after synchronisation the moon's orbital period slowly decreases.  \textit{Middle panel (star--planet locking scenario):} for smaller $M_\mathrm{m}/M_\mathrm{p}$ mass ratio the planet synchronises with the star and becomes tidally locked. \textit{Bottom panel (temporarily locked scenario):} first the planet synchronises with the star (at around $2 \cdot 10^9$ years) but because of the moon's inward migration, later it synchronises with the moon instead ($\sim 7 \cdot 10^9$ years), but after that, the moon shortly reaches the Roche limit of the planet.}\label{massratio}
\end{figure}

\subsubsection{Tidal locking}

In connection with tidal locking, we define three different scenarios. \\

\textit{Moon--planet locking: the Selenoid scenario.}
The evolution track is somewhat similar to the Earth--Moon case. First, the moon migrates outwards, until a spin-orbit synchronisation is reached between the orbital period of the moon and the rotation of the planet. In the later evolution, the moon slowly turns back, the typical time scale before colliding with the planet was in the order of 10 billion years in our integrations. (More specifically, the end state of the moon is reaching the Roche radius of the planet, where the moon will be disrupted). Therefore, we consider this scenario as stable (Fig. \ref{massratio}, top panel). We note that the similarity with the Earth--Moon case is limited to the synchronization between the two bodies, but there is a qualitative difference: here the planet spin is synchronizing with the orbit of the moon, while in the Earth--Moon case the spin of the Moon is synchronized with the orbit of the Earth.

\textit{Star--planet locking.}
This scenario is similar to the Selenoid, but rather than synchronising with the moon, the planet is synchronised to the central star. We did find this possibility in several test runs, and seems to be typical for certain configurations (close-in systems with lightweight moon).  
In these cases, the moon's migration is slow and has a similar time scale than in the Selenoid case. The path is qualitatively identical: first the moon migrates outwards and then later inward (Fig. \ref{massratio}, middle panel). This configuration is also stable on the long term, and the moon can be observed (with difficulties coming from its low mass and small size).

\textit{Temporarily locked scenario.}
We found another tidal evolution scenario with rather complex evolution for which we have not found 
earlier discussion in the literature. 
At the beginning it has essentially the same evolution as the star--planet synchronisation, but since the moon migrates inwards, here the tidal forces from the moon is increasing, and from a point they majorate over the tidal forces from the star. At this point, the planet rotation will be decoupled from the mean motion, and resynchronizes with the moon. The inward migration of the moon is enhanced from that point, and the  moon soon reaches the Roche limit of the planet.
In the scenario shown at the bottom panel of Fig. \ref{massratio}, the planet's synchronisation with the star happens after approximately $2 \cdot 10^9$~years, where there is the turning point of the moon's orbit, too, and the break-up of the star-planet synchronisation occurs at $\sim 7 \cdot 10^9$~years. Because of the large time scales, this configuration is also stable on the long term, and moons in such systems could be found.

\subsection{Time scales}

The time scales of the discussed scenarios depend on the migration rate, hence indirectly on the distance of the Hill-sphere which can limit the evolution path very close to the star. 

Close-in planets always evolve very rapidly, because (i) the Hill-sphere is close to the planet, and the moon reaches the critical distance from the planet very soon, or turns back at a very close distance to the planet; and (ii) since the moon is always very close to the planet, very prominent tidal forces emerge and the orbit evolution will be very fast. For most close-in configurations, similarly to the hottest sub-Jupiters known, the tidal escape time scale of the moon can be even less than 1000 years. The only possible scenarios for those cases when the planet is too close to the star are the disruption and the escape scenarios (Fig. \ref{massratio2} upper and lower panels).

If the planet orbits the star at a larger distance, the path to the the critical distance from the planet is longer and the evolution is slower, hence the time scale of stability can reach billions of years, or even the Hubble time scale (these are the tidal locking scenarios, see Fig.~\ref{massratio}). For selenoids, this means that the fallback takes several billion years, and for runaway moons, the runaway time takes several billion years. In other words, with increasing planet orbital period first we get quickly falling-back selenoids, then stable selenoids (Fig. \ref{massratio2} and \ref{massratio} upper panels).

\section{Known exoplanets} \label{survival}

We calculated the evolution of possible model moons in the known planetary systems with the aim of pre-selecting those exoplanets which can physically have a satellite. These planets might worth a detailed follow-up observation for a moon survey. 
Planets and their model moons were simulated according to the recipe described in Sect. 2. The results are listed in Table \ref{list} (in \ref{table}), ordered by decreasing survival rate. (Only those cases are shown in the table which have non-zero survival rates.) The table also contains the mean of the integration times for each planet. If the survival rate is a hundred percent, then the mean of the integration times will be equal to the age of the star (or of the Sun, if the star's age is not known, as described in Section \ref{Methods}). If in some cases the integration stops earlier, then the mean of the integration times will be shorter. This parameter together with the survival rate can be considered as a measure for the stability of moon orbits.

\begin{figure*}
	\centering   
	\includegraphics[width=37pc]{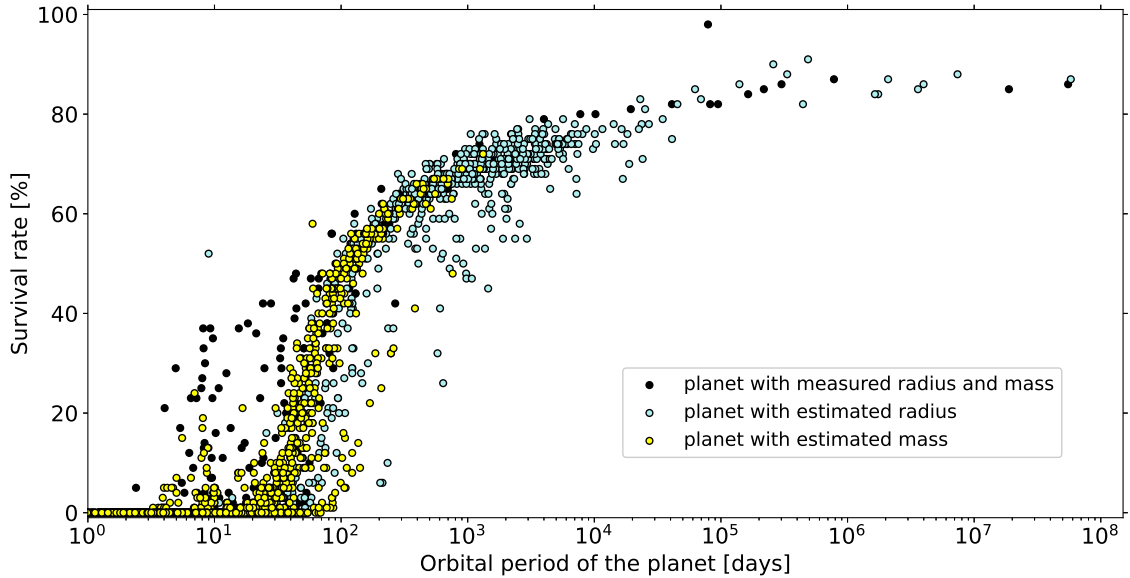}     
	\caption{\label{survival_fig}Survival rate of moons around known exoplanets as function of the planet orbital period. Black coloured dots: planets with known radius and mass data. Teal coloured dots: planets without known radius data, for these, their radius was estimated based on their measured mass (or $M_\mathrm{p} \sin i$). Yellow coloured dots: Planets without known mass data, for these the masses were estimated from their measured radius.
	}
\end{figure*}

Fig. \ref{survival_fig} shows the survival rate for test exomoons as function of the orbital period of the planet. Because the critical escaping distance is small below 10~days orbital period, only a few exomoons stay in a stable orbit around the planet. Above $\sim$~300 days, the survival rate is about 70--90 percent, as a consequence of the large critical distance. Between about 5 and 300~days we find a "transition zone" where the increasing size of the critical escaping distance results in higher and higher survival rates.

The saturation at $\sim$85\% survival rate for large orbital periods is caused by the chosen initial semi-major axes of the moons which in some cases is below the synchronous orbit. In these cases the moons spiral inwards to the planet until they reach the Roche sphere. 

Only a fraction of known exoplanets are promising exomoon hosts according to this work. This is primarily because the orbital period behaves as a moon "censor": not much satellites are expected (survival rate mostly below 40\%) for planets in shorter period orbits than 10 days. 
This is not surprising, since planets this close to the star have small Hill radii, hence stable orbits for a moon are limited to a very narrow space. As the orbit of the moon evolves, sooner or later it will reach the instability limit and escape from the planet's gravitational bond.

Interestingly, the planets with known mass and radius data (see the black dots in Fig. \ref{survival_fig}) are showing smaller deviation above $\sim$100 days orbital period, compared to those for which the radii or masses are estimated from their other parameter (teal and yellow coloured dots, respectively). For the planets with known mass and radius value, the survival rate reaches 50\% at around 100 day orbital period of the planet, while for the other planets it is much more diffused in between approximately a hundred and a thousand days. Choosing a mass or radius value for the planet is an extra free parameter which naturally causes higher uncertainty in the results. This might be the reason of the shift of the \textit{transition zone} from 5--100~days to 30--300~days orbital period for the two kinds of planets. For those where we have to add random mass or radius values, we will find more unstable cases, hence the survival rate starts increasing only at higher orbital periods where the critical escape distance from the planet is larger. Another explanation for the shift could be an observation bias, since the planets for which both mass and radius data are measured are those for which both transit and radial velocity observations were made. This means that they are orbiting close to the star and in the meantime they are also massive. This might also explain why the transition zone is closer to the star for these planets.

For known exoplanets with long orbital periods, the host stars are quite faint, and the S/N ratio required for a successful moon detection is not achievable by even as precise instruments as the Kepler telescope. This finding is the resolution of the seeming paradox of missing exomoons in Kepler data. Moons are more likely to have a stable orbit around long orbital period planets, but these cases are difficult to observe. This result is also an alert for later transit finding surveys to focus on planets around bright stars with $\gtrsim$10--100~days orbital period.

\section{Discussion} \label{discussion}

The successful detection of a moon around an exoplanet relies basically on two independent factors: the moons must exist (occurrence), and those moons must be observable. The occurrence is a product of two factors: the formation rate and the stability time scale. Formation studies of moons (and planets or stars) are beyond the scope of this paper, however, with the described formalism, we estimate the survival rate of moons around currently known planets.

The survival rate of model moons in known exoplanetary systems shows that satellites can only stay in orbit around the host planet if the orbital period of the planet is larger than five days. In our investigations we found a transition zone between 5 and 300 days orbital period where the survival rate is growing with larger orbital periods. Above 300 days orbital period the moon survival rate reaches 70--90\% for the investigated planets. These specific numbers however depend on the range of chosen initial parameters. Also, increasing the integration time could result in decreasing the survival rates in some cases. 
This would be only a quantitative change as it would not alter the relation between the survival rate of moons and the orbital period of the planets. 
Our investigation serves statistical purposes, and it shows a clear trend in survival rates that increase with planetary orbital period.

This result is in good agreement with the findings of \citet{alvarado17}, who developed a more detailed model including the changes of the planet's physical parameters over time as an additional factor influencing the orbital evolution of satellites. Their results showed that moons of planets with lower than $\sim$~60-day orbital periods collide with the planet, but for planets with larger stellar distances, the satellites slowly migrate outwards and stay inside the Hill-sphere of the planet during the whole 4 Gyrs integration time ('realistic' case, Fig. 9 in their article). Including the change in the planet radius and $k_2/Q$ tidal parameter over time in their model increased the chance of moon survival. This implies that neglecting these factors in our study might underestimate the moon survival rates.

Our results are in line with those of \citet{sucerquia19}, too, who concluded that long-period planets are more favourable moon hosts than short-period ones. They used a very similar calculation method with the aim of estimating the TTV and TDV signals for detection. They conclude that in their investigated phase space no system produces TDV signals strong enough to be observed, but there are a very few cases where the TTV signal would be observable with TESS and Kepler. Larger moon--to--planet ratios and stellar distances could help observations.

With a similar aim, but using a different method, \citet{guimaraes18} also investigated known planets which could host moons. Beside checking the possible semimajor-axes of moons, they also took into account their detectability, providing a list of 54 Kepler planets and candidates which could have detectable moons.

Recently, \citet{tokadjian20} made a similar work to ours in which they investigated the orbit evolution of moons around known exoplanets using simplified tidal lag models. They identified 36 habitable zone planets around which a moon might be stable for long time scales (at least for 1~Gyr). We found that from these planets 29 have at least 50\% survival rate for exomoons (typically between 50 and 65\%.). There are also four planets with survival rates higher than 65\%: these are Kepler-459~b (69\%), Kepler-456~b (72\%), Kepler-1654~b (70\%) and Kepler-1647~b (69\%). 
They also found that for 21 planets the time scale before loosing the moon is larger than the Hubble time in all investigated cases. Our results show moon survival rates typically between 55 and 70\% for these planets with mean integration times ($t_\mathrm{mean}$) between 2 and 3~Gyrs (see Table \ref{list}). There is one outstanding case (Kepler-1628~b) which resulted in only 9\% survival rate in our study. The explanation of our lower stability levels in general is that a much bigger phase space was explored for the initial parameters compared to the study by \citet{tokadjian20}. Also, since the age of these planets is typically lower than 5~Gyrs (and in some cases it is not known), our calculation stopped much earlier.

\citet{martinez-rodriguez19} investigated the orbital stability and habitability of exomoons around known planets of M dwarfs. They found 4 planets which were in the habitable zone and for which the orbit of the moon is stable for a long time scale (the migration time of the moon is larger than the Hubble time). These four planets are HIP~12961~b, HIP~57050~b, GJ~876~b and c (in their paper under the names of CD-23~1056~b, Ross 1003~b, IL~Aqr~b and c, respectively). Interestingly, for these planets our model gave 0\%, 0\%, 6\% and 0\% survival rates, respectively. One of these four planets, however, HIP 57050 b was thoroughly investigated also by \citet{trifonov20} (planet name: GJ~1148~b in their paper) where they were using secular theory and direct N-body integrations to study the possibility of a moon. Similarly to our results, they have found that it is unlikely that a moon would stay in stable orbit around the planet.

\citet{zollinger17} investigated the tidal evolution and habitability of exomoons assuming possible configurations, and found that moons in the habitable zone of M dwarfs experience a tidal heating rate comparable to the tidal heat flux of Jupiter's moon, Io ($\sim 2 \mathrm{W/m^2}$). For stars with masses $\lesssim 0.2 M_\odot$, because of the strong gravitational effect of the nearby star, the tidal heating rate of a Mars-like moon around a Jupiter-like planet is much stronger than that of Io's, implying that these moons might not be habitable. The maximum limit of tidal heating for habitability, however, is not known, and atmospheric effects can influence the habitability of bodies, as well \citep{barnes13}.

Recent hydrodynamical simulations showed that icy moons can form from circumplanetary disks of ice giants such as Uranus and Neptune, indicating that there might be a large population of moons around exoplanets beyond the snowline \citep{szulagyi18}. According to our findings, these moons, if formed, can stay in stable orbits for a long period of time.

Older stars host more evolved systems, in which there is a higher possibility that most satellites have disappeared due to either disruption (at the Roche radius of the planet) or escape. The presence of a massive exomoon may prevent synchronization of the planet with the star and promotes synchronization with the moon. In this case, satellites also evolve into a 1:1 spin-orbit resonance, getting to a mutually synchronised planet-moon system. Less dissipative planets (gas giants and ice giants) are more prone to keep satellites because of slower orbital migration \citep[see e.g.][]{sasaki12}. If an exomoon is observed around a planet, it can provide a strong constraint on the value of the quality factor, $Q$, which describes the dissipation in the planet, and gives hint on its interior.

Since younger stars host less evolved systems, there is a higher chance for planets to still have satellites (which will escape or be destroyed later). But satellites on less distant orbits are more difficult to observe \citep{szabo06, simon12}. If more than one moon is found, than their orbital architecture may give information on their formation, and provide a strong constrain on $Q$. It may also be possible to infer the presence of rings.

\section{Conclusion}

There is no confirmed exomoon detection to date, but the quest for finding moons in Kepler data is still in progress. \citet{teachey18a} reported an exomoon candidate, Kepler-1625~b~I \citep[see also][]{teachey18b}, but this discovery is under debate \citep{kreidberg19, teachey19}. Regardless of the existence of this moon, we found stable orbits for satellites around this planet. Recently, the presence of an exomoon candidate was suggested based on observations of sodium and potassium lines around planet WASP-49~b \citep{oza19}. Further observations are needed for confirmation.

As a result of our calculations, we conclude that planets with long orbital periods support high survival rates for moons, and close-in planets, which are easier to observe, are less likely to host moons. This can serve as an explanation for not having a confirmed exomoon discovery so far. 

Thanks to high accuracy photometry and long-term follow-up missions, CHEOPS 
is measuring the radius of numerous known exoplanets through transits with unprecedented accuracy. This paves the way for characterizing the close environment of exoplanets, and can unveil the presence of exomoons and exorings. Moons and rings are natural companions of planets. Neptune sized planets are especially promising targets owing to their (putative) lower dissipation and large mass compared to Earth-like planets. Such targets will be in the reach of the CHEOPS mission. Following the CHEOPS mission, ESA's PLATO 2.0 mission, capable of discovering and following exoplanet systems and host stars, and the ARIEL mission will continue the task of collecting precise exoplanet light curves. Whereas exomoons have never been firmly detected up to date, they should exist and their successful detections will provide invaluable information on the planet's interior and formation process \citep{crida12}. Detecting the first exomoons will be a major discovery as moons could provide habitable environments, in particular if they orbit volatile rich planets (like giant planets) and are located inside, or close to, the habitable zone \citep{forgan16, dobos17}.

\section*{Acknowledgments}
We thank the referee, Alessandro Trani for useful comments which improved the manuscript.

The COFUND project oLife has received funding from the European Union's Horizon 2020 research and innovation programme under grant agreement No 847675. This project has been supported by the Hungarian National Research, Development and Innovation Office (NKFIH) grant GINOP-2.3.2-15-2016-00003, the City of Szombathely under Agreement No. 67.177-21/2016 and the grant KEP-7/2018 of the Hungarian Academy of Sciences. VD has been supported by the Hungarian National Research, Development, and Innovation Office (NKFIH) grants PD-131737 and K-131508. SC acknowledges support for the CHEOPS mission from the CNES agency, and EXOATMOS program from  LabEx UnivEarthS (ANR-10-LABX-0023 and ANR-18-IDEX-0001). 

\newcommand{\newblock}{}
\bibliographystyle{jphysicsB}
\bibliography{ref}


\appendix
\section{Moon survival rate for known exoplanets} \label{table}

See Table \ref{list} for a list of exoplanets in a decreasing order of moon survival rate (see Section \ref{survival} for details). A horizontal line at 50\% moon survival rate separates the cases where moons are more probable to stay in a stable orbit around the planet. Keep in mind that, as discussed in Section \ref{discussion}, the calculations served only statistical purposes, and the actual survival rates listed below might change for different initial conditions and different restrictions.

\onecolumn
{\footnotesize
 \begin{landscape}
\begin{longtable}{lccccccccc}
 \hline\hline
planet name & $M_\star$ [$M_\odot$] & $R_\star$ [$R_\odot$] & $T_\mathrm{eff} [K]$ & age [Gyr] & $M_\mathrm{p}$ [$M_\oplus$] & $R_\mathrm{p}$ [$R_\oplus$] & $P_\mathrm{p}$ [days] & $t_\mathrm{mean}$ [Gyr] & surv. rate [\%] \\
 \hline 
 \endhead 
GQ Lup b & 0.70 & -- & -- & 0.001 & 4.13 & 19.75 & 79025.00 & 0.00098 & 98 \\
HD 163296 c & 2.30 & -- & -- & 0.005 & 95.34 & -- & 487421.45 & 0.00456 & 91 \\
HD 163296 b & 2.30 & -- & -- & 0.005 & 95.34 & -- & 259124.58 & 0.00454 & 90 \\
Fomalhaut b & 1.92 & 1.84 & 8590 & 0.440 & 66.10 & -- & 333592.87 & 0.389 & 88 \\
HD 106906 (AB) b & 1.50 & -- & 6798 & 0.013 & 3495.76 & -- & 7390457.83 & 0.0115 & 88 \\
ROXs 42B (AB) b & 0.60 & -- & 2200 & 0.007 & 2860.17 & 27.43 & 781097.08 & 0.00596 & 87 \\
TYC 8998-760-1 c & 1.00 & -- & 4573 & 0.017 & 1906.78 & -- & 2090805.11 & 0.0146 & 87 \\
WD 0806-661 B b & 0.62 & -- & -- & 1.500 & 2542.37 & -- & 57983276.50 & 1.30 & 87 \\
\end{longtable}
 \end{landscape}
}
\twocolumn

\end{document}